\pdfoutput=1
\documentclass[%
 reprint,
nofootinbib,
 amsmath,amssymb,
 aps,
]{revtex4-2}

\usepackage{amsmath}
\usepackage{tikz}
\usetikzlibrary{quantikz}
\usepackage[caption=false]{subfig}

\usepackage{graphicx}
\usepackage{dcolumn}
\usepackage{bm}
\usepackage{subfiles}
\usepackage[normalem]{ulem}

\usepackage{natbib}
\usepackage{hyperref}

\begin{document}

\preprint{APS/123-QED}

\title{Simulating strongly interacting Hubbard chains with the Variational Hamiltonian Ansatz on a quantum computer
}
\author{Baptiste Anselme Martin$^{1,2}$}%
\author{Pascal Simon$^2$}
\author{Marko J. Ran\v{c}i\'{c}$^1$}%
 \affiliation{1. 
 TotalEnergies, 8, Boulevard Thomas Gobert – B\^{a}t. 861, 91120, Palaiseau}
 \affiliation{2. Universit\'{e} Paris-Saclay, CNRS, Laboratoire de Physique des Solides, 91405, Orsay, France}

\date{\today}

\begin{abstract}

Hybrid quantum-classical algorithms have been proposed to circumvent noise limitations in quantum computers. Such algorithms delegate only a calculation of the expectation value to the quantum computer. Among them, the Variational Quantum Eigensolver (VQE) has been implemented to study molecules and condensed matter systems on small size quantum computers. Condensed matter systems described by the Hubbard model exhibit a rich phase diagram alongside exotic states of matter. In this manuscript, we try to answer the question: how much of the underlying physics of a 1D Hubbard chain is described by a problem-inspired Variational Hamiltonian Ansatz (VHA) in a broad range of parameter values ? We start by probing how much does the solution increases fidelity with increasing ansatz complexity. Our findings suggest that even low fidelity solutions capture energy and number of doubly occupied sites well, while spin-spin correlations are not well captured even when the solution is of high fidelity. Our powerful simulation platform allows us to incorporate a realistic noise model and shows a successful implementation of noise-mitigation strategies - post-selection and the Richardson extrapolation. Finally, we compare our results with an experimental realization of the algorithm on IBM Quantum's \texttt{ibmq\_quito} device.

\end{abstract}

\maketitle

\section{Introduction}\label{sec:intro}

Quantum computers are a potentially disruptive technology \cite{feynman2018simulating, deutsch1992rapid, shor1999polynomial, grover1996fast} promising chemical precision solutions of problems where electron-electron correlations play a major role. A number of algorithms have been proposed for both the currently non-existing quantum error-corrected devices \cite{PhysRevLett.79.2586, PhysRevLett.83.5162} and the currently available NISQ devices \cite{preskill2018quantum}. The Variational Quantum Eigensolver (VQE) \cite{peruzzo} and Quantum Imaginary Time Evolution (QITE) \cite{QITE, mcardle2019variational} are two NISQ front runners for solving problems in chemistry and condensed matter physics.

The Hubbard model \cite{hubbard1963electron} is one of the most studied problems in condensed matter physics. Elegant and seemingly simple it has been the starting point in studying high-temperature superconductivity \cite{lee2007high} and metal-to-insulator transitions \cite{mott1949basis}, just to name a few. Although the Hubbard model is analytically solvable in one dimension with the famous Bethe ansatz \cite{bethe1931theorie}, a solution in two and three dimensions is generally unknown and has been the focus of half of century of numerical studies \cite{scalapino2007numerical, leblanc2015solutions}.

In this manuscript, we use an 8-site 1D Hubbard model as a benchmarking tool for contemporary quantum computing algorithms, or to be more precise the VQE with the Variational Hamiltonian Ansatz (VHA) \cite{vha, vhagateerror}. The VHA leverages an ansatz that is inspired by the Hubbard Hamiltonian itself and an adiabatic way of thinking to construct the trial wavefunction. Adopting a condensed matter physics perspective, we try to quantify how much of the underlying physics is preserved with an ansatz that minimizes the energy and respects all physical symmetries of the system. Our advanced simulation platform allows us to incorporate a realistic noise model, to quantify how much is the VHA influenced by pure dephasing, spin relaxation, and gate errors and to compare with experimental results from the IBM Quantum's \texttt{ibmq\_quito} device. Furthermore, we attempt an error mitigation strategy called the Richardson extrapolation  \cite{error_mitig, kandala} - a measurement of an observable with artificially modulated noise levels followed by an extrapolation to the zero-noise case, alongside with post-selection.

The adiabatic evolution inspired VHA was first introduced in Ref. \cite{vha} and have been investigated for Hubbard models in Refs. \cite{vhagateerror, strat_cade, resource_estim}. Furthermore, studies of spin systems were conducted in Ref. \cite{entang_hva} as well as optimization studies with the well-studied Quantum Approximate Optimization Algorithm (QAOA) which share the same structure (proposed before VHA in Ref. \cite{qaoa}). The study conducted here advances that of Ref. \cite{vhagateerror} with a more elaborate noise model and discussions of potential noise-mitigation strategies. Furthermore, spin correlations, noise mitigation, and a broader range of $U$ are discussed as compared to Ref. \cite{strat_cade} and focuses on how well certain variables are captured as compared to resource estimation as in Ref. \cite{resource_estim}.

Our findings suggest that different quantities of the physical system require different levels of ansatz complexities. Namely, even for an ansatz of relatively low depth doubly occupied site number and energy are predicted at a satisfactory level. On the other hand, long-range spin correlations are much more sensitive to the infidelities of the trial solution - even a 90$\%$ fidelity is insufficient to capture the correct long-range spin-spin behaviour. Finally, Richardson extrapolation strongly mitigates the effects of quantum noise, although we are only able to validate this for a 2-site Hubbard model (4 qubit simulation).

This paper is organized as follows, in Section \ref{sec:Methods} we describe the methodology, in \ref{sec:Res} we show noiseless results for our estimation of energy, fidelity, doubly occupied site number and spin-spin correlations. In Section \ref{sec:noise} we discuss the noise model, post-selection, the Richardson extrapolation and experimental results before concluding in Section \ref{sec:conclu}.

\section{Methods}\label{sec:Methods}
Variational Quantum Eigensolver (VQE) has been increasingly popular in quantum computing.  First proposed and implemented in Ref. \cite{peruzzo}, VQE offers a method to obtain approximate groundstates of electronic systems described by an Hamiltonian $H$. The VQE algorithm commonly has the following structure:
\begin{itemize}
    \item [1.] An initial state $|\Psi_0\rangle$ is prepared on the quantum computer.
    \item [2.] Subsequently, one applies a variational circuit $\mathcal{U}(\boldsymbol{\theta})$, dependent on a set of variational parameters $\boldsymbol{\theta}$, yielding a variational state $|\Psi(\boldsymbol{\theta})\rangle = \mathcal{U}(\boldsymbol{\theta})|\Psi_0\rangle$.
    \item [3.] After choosing a set of initial parameters $\boldsymbol{\theta}_0$, we prepare $|\Psi(\boldsymbol{\theta}_0)\rangle $ and measure the corresponding energy $E(\boldsymbol{\theta}_0) = \langle \Psi(\boldsymbol{\theta}_0)|H|\Psi(\boldsymbol{\theta}_0)\rangle $.
    \item [4.] Based on the energy, a classical minimizer will optimized the parameters $\boldsymbol{\theta}$ until it converges to a minimal energy.
\end{itemize} 
VQE is particularly suited to NISQ devices as it allows to obtain results with short circuits but also to mitigate errors thanks to the flexibility of the variational parameters.

The choice of the parametrized circuit is essential in the VQE procedure. Two key elements need to be taken into account when selecting the type of the parametrized circuit: the ability to produce the groundstate with sufficiently high fidelity, and the implementability of the circuit on contemporary hardware. These two requirements are intertwined due to the limited length, connectivity and qubit number current-day quantum circuits can have, but also the number of variational parameters a classical optimizer can handle. We can usually distinguish between two strategies when designing an ansatz: 1) the so-called hardware efficient ansatz prioritizes on two-qubit gates which can be naturally performed in the system due to a specific qubit topology \cite{he, Express_entang_he}. Although providing satisfying results on problems that require a small number of qubits, this approach often fails to scale to larger problems and can suffer from optimization barren plateaus \cite{barren}. 2) Numerous approaches proposed problem-inspired ansätze which produce variational states from relevant subspaces \cite{UCC_vqe, lcda, local_exp, Agate, Symmetry_adapt, simulating_static}. Built from physical arguments, they often require more costly circuits. Furthermore, adaptive schemes have been proposed to reduce the circuit length of those methods \cite{adapt,qubit_adapt}.

Let us now present the thinking behind constructing a VHA trial solution inspired in the Hamiltonian of the problem itself. Observe two Hamiltonians, $H_0$ and $H_f$. $H_0$ corresponds to an eigenproblem which is easily solvable and $H_f$ to a problem which is difficult to solve. The basic idea in adiabatic evolution is to prepare the groundstate $|\Psi_0\rangle$ corresponding to the initial Hamiltonian $H_0$.  Then, one can evolve the initial state in time $T$ described by the following time-dependent Hamiltonian $H(s) = (1-s)H_0 + sH_f$, with $s(t=0) = 0$ and $s(t = T) = 1$. If the evolution rate is slow enough with respect to the gap of the system, Fock's adiabatic theorem  guarantees that $|\Psi(t)\rangle = \mathcal{T}e^{-i\int_0^t H(s(\tau))d\tau} |\Psi_0\rangle $ (where $\mathcal{T}$ is the time-ordering operator) remains in the groundstate of the instantaneous Hamiltonian $H(s)$.

In general, and on a quantum computer, the time-evolution operator can be approximated with a first order Trotter-Suzuki approximation by dividing the integration over time into $N$ time step of duration $\Delta\tau$ so that $T = N\Delta\tau$: $\mathcal{T}e^{-i\int_0^T H(s(\tau))d\tau} = \prod_{n=1}^N e^{-iH(s(n\Delta\tau))\Delta\tau} \simeq  \prod_{n=1}^N e^{-i(H_0(1-s(n\Delta\tau)) + H_f s(n\Delta\tau))\Delta\tau} $. Secondly, we decompose $H$ into non-commuting parts $\{H^\alpha\}$: $H = \sum_\alpha H^\alpha$. The terms contained in $H^\alpha$ should commute among each other in order to implement them simultaneously. By using again the Trotter-Suziki approximation, we obtained the following circuit:

\begin{multline}
    \mathcal{T}e^{-i\int_0^T H(s(\tau))d\tau} \simeq \\ \prod_{n=1}^N\prod_\alpha e^{-iH_0^\alpha(1- s(n\Delta\tau))\Delta \tau} \prod_\beta e^{-iH_f^\beta s(n\Delta \tau) \Delta \tau}.
\end{multline}
However, to respect the adiabatic criterion long evolution times are required leading to long circuits beyond the reach of NISQ devices. In order to shortcut adiabatic paths, VHA replaces the time steps by variational parameters, with the hope of reducing the circuit depth needed to obtain satisfying states. The ansatz can be formulated by the following:
\begin{equation}
     \mathcal{U}(\boldsymbol{\theta}) = \prod_{n=1}^{N_L}\prod_\alpha e^{-iH_0^\alpha \theta_n^\alpha} \prod_\beta e^{-iH_f^\beta \theta_n^\beta},
\end{equation}
where, $N_L$ is the number of layers, and $\boldsymbol{\theta}$ are the variational parameters.
\def\myvdots{\ \vdots\ }

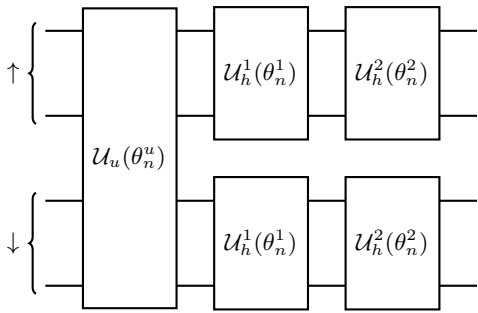
\begin{figure}
     \centering
       \begin{quantikz}
&\lstick[wires=2]{$\uparrow$}&\gate[wires=4]{\mathcal{U}_u(\theta_n^u)}&\gate[wires=2]{\mathcal{U}_h^1(\theta_n^1)}&\gate[wires=2]{\mathcal{U}_h^2(\theta_n^2)}&\qw\\
&&                 &                   &                   &\qw\\
&\lstick[wires=2]{$\downarrow$}&                &\gate[wires=2]{\mathcal{U}_h^1(\theta_n^1)}&\gate[wires=2]{\mathcal{U}_h^2(\theta_n^2)}&\qw\\
&&                 &                   &                   &\qw\\
\end{quantikz}  
     \caption{One layer for VHA, where $\mathcal{U}_u(\theta)= e^{-iH_U \theta}$, $\mathcal{U}_h^1(\theta) =  e^{-iH_t^{(1)} \theta} $, and $\mathcal{U}_h^2(\theta) =  e^{-iH_t^{(2)}\theta}$. } 
     \label{fig:schematic_circuit}
 \end{figure}

Let us consider now the case of the 1D Hubbard model, described by the following Hamiltonian
\begin{multline}
    H=-t\sum_{i\sigma} \left(c^\dagger_{i\sigma} c_{i+1\sigma}+ \text{h.c.}\right)+\\
    U\sum_i \left(n_{i\uparrow}-\frac{1}{2}\right)\left(n_{i\downarrow}-\frac{1}{2}\right).
\end{multline}
For the Hubbard Hamiltonian and in the context of trying to determine its groundstate with adiabatic methods one can choose $H_0$ to be the free fermion Hamiltonian $H_t = -t\sum_{i\sigma} (c^\dagger_{i\sigma} c_{i+1\sigma}+ c^\dagger_{i+1\sigma} c_{i\sigma}) $ and the final Hamiltonian to be the full Hubbard Hamiltonian, which corresponds to the situation of slowly increasing the interaction strength $U$ in the system. While a one-body state can be in principle efficiently prepared on a quantum computer \cite{solving_strongly_corr, ZhangJian, openfermion}, the variational circuit can be formulated by:
 \begin{equation}\label{eq:vha_hubbard}
      \mathcal{U}(\boldsymbol{\theta})= \prod_{n=1}^{N_L}  e^{-iH_t^{(2)}\theta_n^2}e^{-iH_t^{(1)} \theta_n^1}e^{-iH_U \theta_n^U}.
 \end{equation}
 with $H_U=U\sum_i \left(n_{i\uparrow}-\frac{1}{2}\right)\left(n_{i\downarrow}-\frac{1}{2}\right)$, $H_t^{(1)} = \sum_{i \in \text{even},\sigma} \left(c^\dagger_{i\sigma} c_{i+1\sigma}+ \text{h.c.}\right)$, $H_t^{(2)} = \sum_{i \in \text{odd},\sigma} \left(c^\dagger_{i\sigma} c_{i+1\sigma}+ \text{h.c.}\right)$. Fig. \ref{fig:schematic_circuit} shows a schematic of one layer of the ansatz.
 
\section{Results}\label{results}

\label{sec:Res}
\begin{figure}
     \centering
     \includegraphics[width = 0.9
     \linewidth]{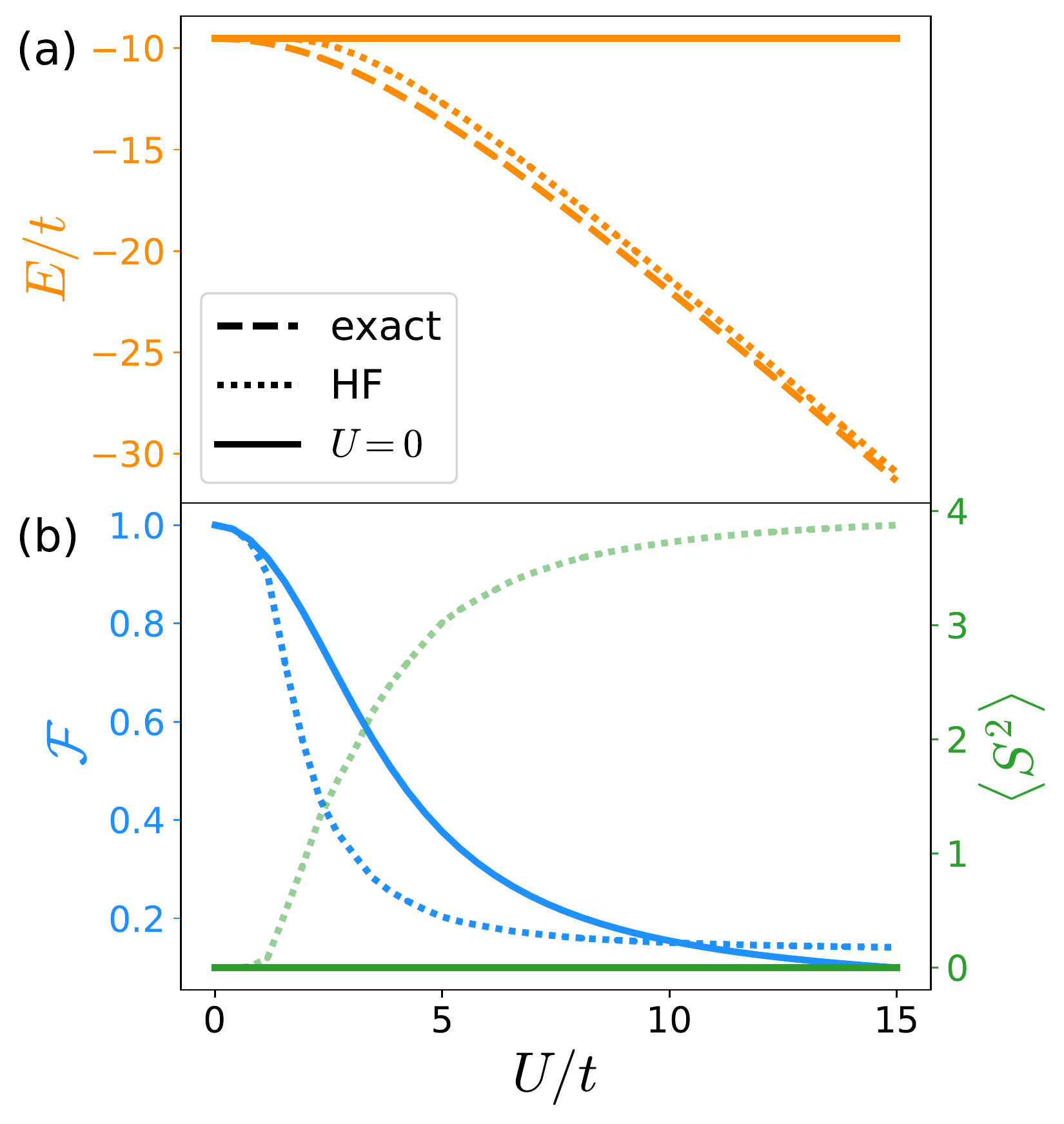}
     \caption{Comparison of $U=0$ free fermion and mean-field Hartree Fock state as the initial state. (a) The energy $\langle \Psi |H(t,U)| \Psi\rangle$ and (b) the fidelity and the total spin $\langle S^2 \rangle$ value as a function of $U$ for a $N = 8$ chain. The fidelity of the exact solution is trivially one. For both (a) and (b), the full line, dashed line and dotted line designate respectively the $U=0$, mean-field Hartree Fock, and exact solution.}
     \label{fig:initial_state_comp}
 \end{figure}

 We start by comparing different ways of preparing the initial state of the Hubbard model on a quantum computer. We will restrict ourselves to one-body state as initial states due to the simplicity of their preparation. We investigate two natural choices: a free fermion $U=0$ groundstate and a mean-field Hartree-Fock (HF) solution as the latter is commonly utilized in VQE studies in the chemistry domain with widely studied Unitary Coupled Cluster Ansatz (UCC) \cite{originalUCC, UCC, UCC_vqe} (recently schemes to prepare Gutzwiller wave function was also proposed in Refs. \cite{gutzwiller, seki2022gutzwiller}). In similarity to problems in chemistry the HF state provides a good approximation to the groundstate energy (see Fig. \ref{fig:initial_state_comp}a). 
 
 However, often in the study of condensed matter systems, one is interested in more quantities than just the mere energy such as long-range correlations or the expectation value of the square of the total spin $\langle S^2 \rangle$. The most comprehensive measure of the quality of the obtained state is the fidelity $\mathcal{F} = |\langle \Psi(\boldsymbol{\theta})|\Psi_{\rm ex}\rangle|^2$, where $\Psi(\boldsymbol{\theta})$ is the approximate solution and $\Psi_{\rm ex}$ the exact solution. The fidelity will be $0$ if the approximated solution is totally different from the exact one and $\text{1}$ if these two solutions match.
 
The mean-field solution is obtained by transforming the two-body term in the Hamiltonian responsible for interaction into a one-body term thanks to the following approximation: $\sum_i n_{i\uparrow} n_{i\downarrow} \simeq \sum_i \left(\langle n_{i\uparrow} \rangle n_{i\downarrow} + \langle n_{i\downarrow} \rangle n_{i\uparrow} \right)$.  Starting from initial guess $\{\langle n_{i\sigma} \rangle_0\}$, the average densities $ \{\langle n_{i\sigma} \rangle \}$ are tuned self-consistently to minimize the energy $E_\text{HF} = \langle \Psi_\text{HF} | H_{\text{HF}}(\{\langle n_{i\sigma}\rangle \})|\Psi_\text{HF}\rangle $ where $\ket{\Psi_\text{HF}}$ is the groundstate of the mean-field Hamiltonian $H_\text{HF}(\{\langle n_{i\sigma}\rangle\}) = H_t + U\sum_i \left(\langle n_{i\uparrow} \rangle n_{i\downarrow} + \langle n_{i\downarrow} \rangle n_{i\uparrow} \right) -  \frac{U}{2} \sum_{i\sigma} n_{i\sigma} $. The total spin operator $S^2$  can be express as $S^2 = (S^x)^2 + (S^y)^2 + (S^z)^2$ with $S^\alpha = \sum_i S^\alpha_i$ with $\alpha =x,y,z$ and $i$ the position index. One can first express those operators in terms of fermionic operators:  $S^x_i = \frac{1}{2}(S^+_i + S^-_i) $, $S^y_i = \frac{1}{2i}(S^+_i - S^-_i) $ , $S^z_i = \frac{1}{2}\left(n_{i\uparrow} - n_{i\downarrow}\right)$ with the spin raising and lowering operators $S^+_i = c^\dagger_{i\uparrow}c_{i\downarrow}$ and $S^-_i = c^\dagger_{i\downarrow}c_{i\uparrow}$. Thanks to the chosen fermion-to-qubit mapping described in the Appendix \ref{subsec:qb_gate}, one can write those fermionic operators into qubit operators which can in principle be measured on a quantum hardware.  
 
In terms of fidelity, the free fermion solution is much closer to the exact solution in the regime of $0\le U/t \approx 10$ Fig. \ref{fig:initial_state_comp}b. Moreover, the free fermion solution respects the $S^2$ symmetry of the problem. On the other hand, the Hartree-Fock solution produces non-zero values for the total spin $S^2$. Note that a Restricted Hartree-Fock calculation will return the free fermion groundstate as the Hubbard model contains only two-body interactions between opposite spin electrons from the same site (i.e. same spatial orbital). As VHA is Hamiltonian-based, the symmetries are preserved throughout the circuit (all the gates commute with the particle number operator $N$ and the spin operators $S^z$ and $S^2$). Therefore choosing an initial state with a right set of quantum numbers avoids the variational ansatz to span non-physically relevant parts of the Hilbert space. Other strategies aiming at including symmetries in the VQE process have been also proposed in \cite{penalty, Symmetry_adapt, DenisLacroix, yen2019exact, seki2021spatial}.

We use an iterative approach consisting in initializing our simulations by targeting an initial value of Coulomb repulsion from the intermediate regime: $U_0 = 5$ (we choose $t=1$). We try many random initial parameters contained in $[-\pi/5,\pi/5]$, keep the lowest energy solution, and use the optimized parameters as initial parameters to simulate the case $U_1 = U_0 \pm \delta U$. Throughout the simulations, the optimized parameters for the case $U_{N-1} = U_0 \pm (N-1)\delta U$ are used as initial parameters for $U_N = U_{N-1} \pm \delta U$. This method is motivated by the fact that the choice of initial parameters has a significant impact on the optimized solution (as detailed in the Appendix \ref{subsec:opt_details}), and by trying multiple starting parameters, we mitigate the effect of initial parameters choice, which improves the result for the whole range of $U$ values and speed up the optimization routine. 
\begin{figure}[!htb]
    \centering
    \includegraphics[width=0.8\linewidth]{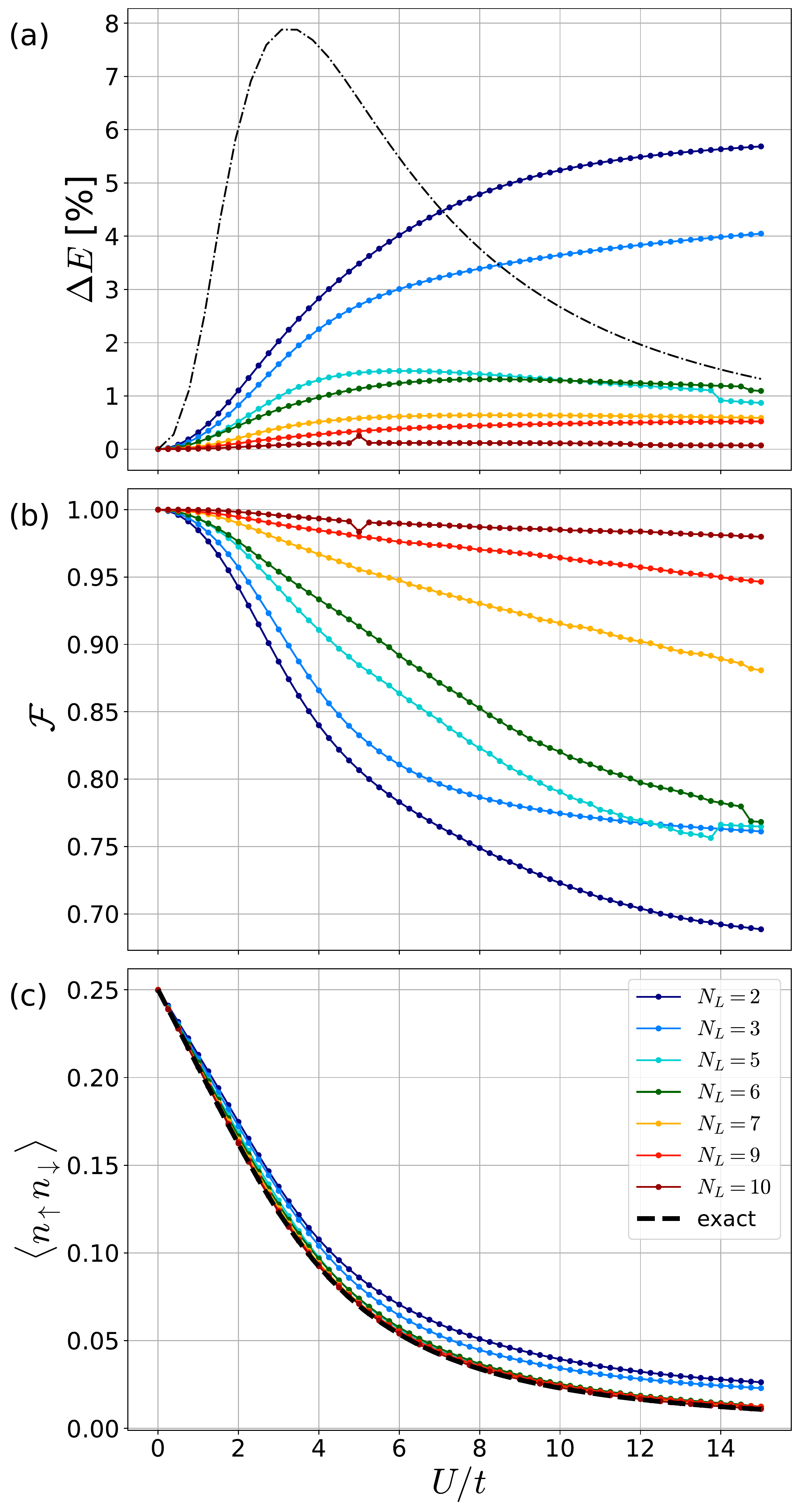} 
    \caption{Classical simulation of the VQE algorithm applied to the 1D Hubbard model with 8 sites (16 qubits), with $U_0 = 5$. (a) The energy error at different circuit depths. (b) The fidelity of the output state with the exact groundstate. (c) The average number of doubly occupied sites $\langle n_{\uparrow} n_{\downarrow}\rangle$.}
    \label{fig:E_fid_Lx8}
\end{figure}

\begin{figure}[htb]
    \centering
    \includegraphics[width=0.8\linewidth]{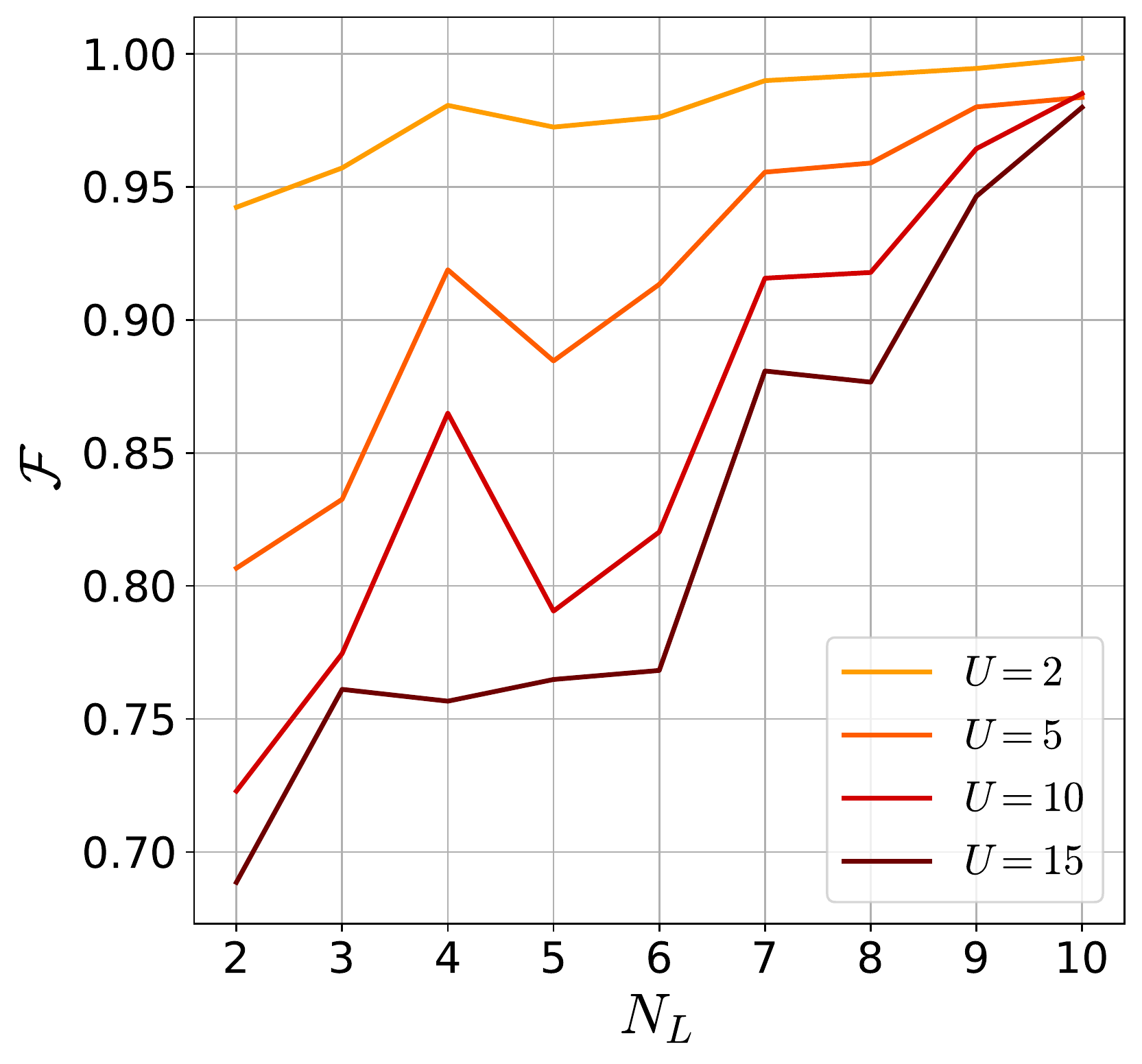}
    \caption{The fidelity of the optimized states with exact solution as a function of the number of layers $N_L$ for different values of $U$. }
    \label{fig:fid_vs_NL_vs_U}
\end{figure}

Fig. \ref{fig:E_fid_Lx8} shows the results for a $N = 8$ Hubbard chain. Whereas energy and fidelity errors remain small in weakly interacting regime $U \simeq 2$ even for limited numbers of ansatz layers $N_L$, the non-interacting initial state already providing a close match, quality of results decreases as $U$ increases. This can be understood as the result of lower fidelity of the initial state (Fig. \ref{fig:initial_state_comp}). Fig. \ref{fig:E_fid_Lx8} shows the effect of the number of layers $N_L$ on the fidelity of the optimized states at different values of $U$. To simulate a Hubbard chain in a low interacting regime, only a few layers are required to reach relatively high fidelities $> 95\%$. However, as $U$ increases, the overlap between the initial and the exact solution becomes smaller. Adding more layers to the parametrized circuit leads to better results, although not systematically - highlighting the complex nature of the resulting variational state. This is shown for example that the $N_L=3$ ansatz produces higher energy states than the $N_L = 5$ case but leads to similar fidelity results in the region $U \gtrsim 12$. This shows that the ansatz produces polluting states that have low energies but are further away from exact solutions. Similarly, Fig. \ref{fig:fid_vs_NL_vs_U} shows that $N_L = 4$ ansatz produces better fidelities than $N_L = 5,6$ simulations. One can note a kink in the $N_L = 10$ results at $U = 5$. This shows that the multi-start parameters still resulted in a false minima value with slightly higher energy and lower fidelity. However, solutions were improved when simulating the next case $U\pm\delta U$. 

Despite the fact that low depth variational circuits fail to produce the exact groundstate of the 1D Hubbard chain, we can still wonder how good are those approximate solutions describing some particular expected behaviour of our system. Fig. \ref{fig:E_fid_Lx8}b shows the fidelity as a function for $U$ for different depths and its corresponding values for $\langle n_{\uparrow} n_{\downarrow}\rangle = \frac{1}{N}\sum_i \langle n_{i\uparrow} n_{i\downarrow}\rangle $, $N$ being the number of sites. Despite having a low overlap with the exact solution, the optimized states recover pretty well the transition from delocalized to localized electrons as the on-site Coulomb repulsion increases. This shows that the ansatz builds up non-trivial correlations to a non-interacting state, and captures this basic manybody phenomenon.
\begin{figure*}
    \centering
    \includegraphics[width = 0.8\linewidth]{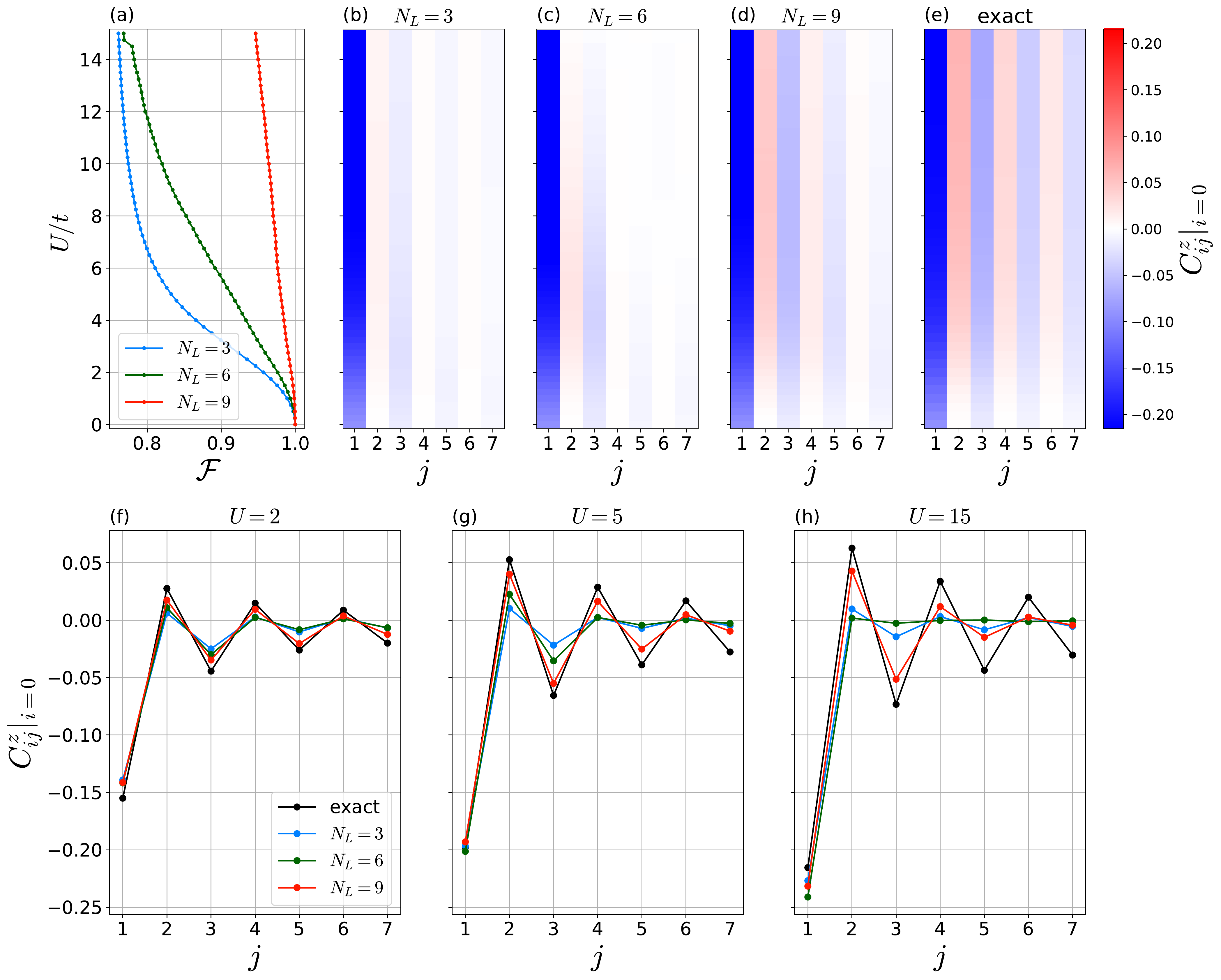}
    \caption{(a) The fidelity $\mathcal{F}$ as a function of $U/t$ for $N_L = 3,6$ and $9$ VQE solutions (8-site Hubbard chain). (b,c,d,e) Spin correlation $C^z_{ij} = \langle S^z_i S^z_j \rangle - \langle S^z_i\rangle \langle S^z_j \rangle$ calculated between one end of the chain ($i=0$) and the other sites ($j\geq 1$) as a function of $U$ for $N_L = 3,6,9$ and exact solution. (f,g,h)  $ C^z_{ij} = \langle S^z_i S^z_j \rangle - \langle S^z_i\rangle \langle S^z_j \rangle$ for $i=0$ as a function of $j\geq 1$ for $U=2,5$ and $15$ for a number of ansatz layers $N_L = 3,6,9$.}
    \label{fig:CSz}
\end{figure*}


\
Secondly, we take a look at long-range spin correlations in our variational states, defined by $ C^z_{ij} = \langle S^z_i S^z_j \rangle - \langle S^z_i\rangle \langle S^z_j \rangle$. Because two electrons of opposite spin will repel each other on the same site, increasing $U$ will induce an antiferromagnetic order in the chain. As a result, we expect that the spin value of one end of the chain is correlated with the sites along the chain. This is shown in Fig. \ref{fig:CSz} where we get sign-alternating values for $C^z_{ij}|_{i=0}$. We observe that as $U$ increases, the ansatz fails to capture long-range correlations.


\section{Noisy Simulations and experiment}

\label{sec:noise}
VQE algorithms were introduced to extend the capabilities of current hardwares, strongly limited by noise and decoherence. A natural question when investigating the performance of an ansatz is how robust it is when noise occurs in the hardware. Here we investigate how noise degrades the output of the algorithm. To do so, we include idle noise and gate errors in our simulation. Idle noise will occur on inactive qubits that undergo spontaneous relaxation from $\ket{1}$ to $\ket{0}$ and loss of quantum information (ie loss of the phase difference knowledge) between $\ket{0}$ and $\ket{1}$ states (thus making the qubit more and more "classical"). The processes will be characterized by the typical timescales $T_1$ and $T_2$. Gate errors will be modelled by a depolarizing channel motivated from randomized benchmarkings \cite{depol_rand} which assume a probability of error $p_1$ ($p_2$) for single-qubit (two-qubit) gates. The quantum channels used to model noise are detailed in the Appendix \ref{subsec:noise_model}.
 
To set the inactive duration, we assign an average duration of $t_1 = 60$ ns for single-qubit gates and $t_2 = 425$ ns, as well as $T_1 = 120$ $\mu$s and $T_2 = 85$ $\mu$s. For gate errors, we set $p_1 = 0.3 \%$ and $p_2 = 1 \%$  which corresponds to current-day quantum computers performances from IBM \cite{IBM}.

\begin{figure*}
    \centering
    \includegraphics[width=0.8\linewidth]{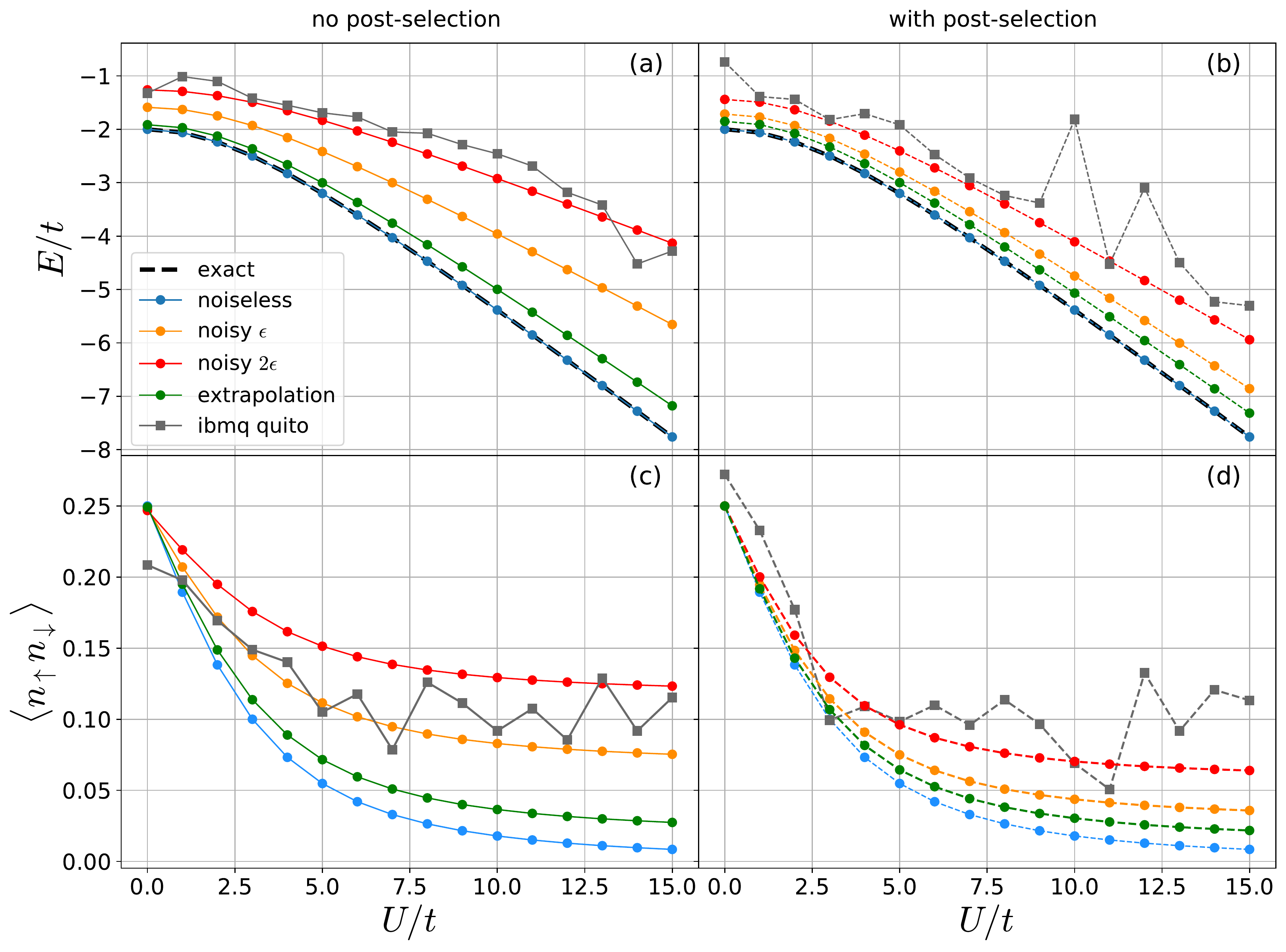}  
    \caption{Noisy simulation with two level of noise, $\epsilon_1 = 3.10^{-3}$ and $\epsilon_2 = 10^{-2}$, and $2\epsilon_1, 2\epsilon_2$ and experimental data from IBM Quantum's \texttt{ibmq\_quito}. Panels (a) and (c) ((b) and (d)) show the energy and the double occupancies as a function of $U$ without (with) post-selection.}.
    \label{fig:rich_ex}
\end{figure*}

The simulations were based on density matrix calculations. Fig. \ref{fig:rich_ex} shows the obtained results for 2-site Hubbard system and simulations incorporating noise being compared with noiseless simulations. Whereas for the noiseless case exact solutions are obtained with only one layer of ansatz for all value of $U$, noise affects greatly the quality of the results. While idle noise induces an infidelity of about 2\%, adding gate errors degrade significantly the results, indicating that it is the major source of error.

In an attempt of mitigating those errors, we test two error mitigation techniques: post-selection and zero-noise extrapolation. As described in Appendix \ref{subsec:postselec}, post-selection uses the \textit{a priori} known symmetries of the target solution to discard any measurement output that doesn't respect those physical constraints. In this work, we rejected the shots not respecting the number of particules and $S^z$ symmetries. To further improve results, we simulate the same circuit with error rate $p_2$ and $2p_2$ and perform Richardson extrapolation as proposed by \cite{error_mitig} and experimentally implemented in \cite{kandala}. The idea is to increase on purpose the noise level on quantum hardware. An observable quantity $\langle \hat{O}\rangle$ can be written as $\langle \hat{O}\rangle = O^* + \sum_k c_k \epsilon^k$, $\epsilon$ being the quantity representing to noise level and $O^*$ the targeted noiseless value (see Appendix \ref{subsec:richardson}). Therefore, by doubling the error rate of the gates, one can perform a first-order extrapolation. This gives an approximation of the noiseless value of the energy $E^*$ at each iteration of the minimization process, but also for other quantities like $\langle n_\uparrow n_\downarrow\rangle$ measured from the optimized state. Fig. \ref{fig:rich_ex} shows that both techniques can significantly improve the results in terms of energy and number of doubly occupied site values.

Finally, we test those results against real implementation of the algorithm on IBM's hardware \texttt{ibmq\_quito} \cite{IBM} as shown in Fig. \ref{fig:rich_ex}. Although the benchmark parameters were close to the noise parameters ($\epsilon_1^{\text{exp}} \simeq 2.6.10^{-3}$, $\epsilon_2^{\text{exp}} \simeq 1.2.10^{-2}$ $T_1^{\text{exp}} = 96 \mu s$ and  $T_2^{\text{exp}} = 102 \mu s$) at the time of the experiment, the experimental results seem more affected by noise. This can be explained by the absence of read-out and shot noise in our model (the experiment's number of shots was set to 20,000), but also by the fact that additional noise occurs in the hardware like crosstalk noise. Finally, additional SWAP gates (decomposed with 3 CNOTs) are needed because of the hardware's topology. However, the experimental results are in qualitative agreement with the noisy simulations as both energy and double occupancies follow the same behaviour. While we didn't implement zero-noise extrapolation as it represents a difficult experimental challenge, post-selection improves overall the results for the energy values as well as the number of doubly occupied sites, where the steep decrease for small values of $U$ is better captured.

\section{Conclusion}\label{sec:conclu}

We investigated the Variational Hamiltonian Ansatz (VHA) \cite{vha} through the example of the one-dimensional Hubbard model. This ansatz is inspired by the natural idea of adiabatic evolution, in which a free fermion initial state is driven towards the correlated groundstate by slowly increasing interaction in time. Each time steps are replaced by variational parameters in the hope to find an optimized path to the groundstate. We perform classical simulations of the VQE algorithm for an 8-site Hubbard chain which corresponds to 16 qubit simulations. We have also carried out simulations up to 12 sites (24 qubits) which are giving qualitatively equivalent results. However, we have chosen to present the 8-site chain results as it represents a good trade-off between system size and computational time. 
  
By calculating the energy error and the fidelity of the optimized states, we quantify the performances of the algorithm. Although low numbers of layers give satisfying results in weakly correlated regimes, higher circuit depths are required to target strongly interacting groundstates. We study how well the approximated groundstates capture the physics of the Hubbard model by investigating correlations built upon the free fermion Slater determinant. We observe that short-length circuits can capture well the localization of electrons as $U$ increases. Long-range antiferromagnetic correlations are however harder to obtain despite high fidelity states and require a larger number of ansatz layers.

Finally, we test the ansatz against noise for a 2-site Hubbard model. We include noise models in our simulations based on density matrix calculations and compare with experimental results from IBM Quantum's \texttt{ibmq\_quito} device. Because of two-qubit gate errors results are greatly degraded, which indicates the practical limit of Trotterized-like ans\"{a}tze. Zero-noise extrapolation enables us to qualitatively improve our noisy simulation results for both the energy and the number of doubly occupied sites. Further improvements should be aimed at both proposing relevant initial states requiring short circuits still exhibiting the right symmetries as well as bridging this scheme towards hardware efficient approaches.

\section*{Acknowledgment}
We are grateful to the Association Nationale de la Recherche et de la Technologie for funding. The computations were carried out on the Atos Quantum Learning
Machine (QLM). We acknowledge IBM Quantum for the access of their quantum devices \cite{IBM}. The quantum circuit schematics were drawn thanks to Quantikz package \cite{kay2018tutorial}.\\
\textit{Note added:} After submitting our manuscript, an independent preprint \cite{stanisic2021observing} investigated experimentally spin and charge properties of the Hubbard model by means of a 1-layer VHA circuit. The reported results are in qualitative agreement with our study.

\section*{Appendix}\label{SI}

 \subsection{Qubit mapping and gate decomposition}\label{subsec:qb_gate}

In order to express the ansatz in terms of a qubit circuit, one needs to map the fermionic operators onto qubit operators. Among the different existing mappings, we choose the Jordan-Wigner transformation. To implement this, one first needs to choose an ordering for the different one-body states. For a Hubbard chain of length $N$, we choose to index the states by site index, the $N$ first being attributed to spin-up, the $N$ second for spin-down states, so that $ c^\dagger_{i\uparrow} = c^\dagger_{i}$ and $ c^\dagger_{i\downarrow} = c^\dagger_{i + N}$. We express the creation and annihilation operators as follows:

\begin{equation}
    c^\dagger_{i} = \otimes_{j<i} Z_j \otimes  \begin{pmatrix} 0&1\\
    0&0\end{pmatrix}  = \otimes_{j<i} Z_j \otimes \frac{X_i - iY_i}{2}
\end{equation}
\begin{equation}
    c_{i} = \otimes_{j<i} Z_j \otimes  \begin{pmatrix} 0&0\\
    1&0\end{pmatrix}  = \otimes_{j<i} Z_j \otimes \frac{X_i + iY_i}{2}
\end{equation}
Having express fermionic operators into Pauli strings, one can rewrite the operator contained in the ansatz:

\begin{multline}\label{eq:hopping}
    e^{-i(c^\dagger_i c_{i+1} + c^\dagger_{i+1} c_i)\theta} =  e^{-i(X_i X_{i+1} + Y_i Y_{i+1})\theta/2}\\
    = \begin{pmatrix} 1&0&0&0\\ 0&\cos(\theta)& -i\sin(\theta)&0\\
    0&-i\sin(\theta)& \cos(\theta)&0\\
    0&0&0&1
    \end{pmatrix},
\end{multline}

\begin{multline}\label{eq:onsite}
    e^{-i(n_i- \frac{1}{2})(n_{i + N} - \frac{1}{2})\phi} =  e^{-iZ_i Z_{i+N}\phi/4} \\
    = \begin{pmatrix} e^{-i\phi/4} &0&0&0\\ 0&e^{i\phi/4}& 0&0\\
    0&0& e^{i\phi/4}&0\\
    0&0&0&e^{-i\phi/4}
    \end{pmatrix}.
\end{multline}

In order to prepare one-body states as initial states, we use a method based on QR decomposition \cite{solving_strongly_corr, ZhangJian} implemented in the OpenFermion package \cite{openfermion} that decompose a basis transformation into a set of Givens rotations, that can be expressed as:
\begin{multline}\label{eq:givens}
    G(\varphi) = e^{i(c^\dagger_i c_{i+1} - c^\dagger_{i+1} c_{i})\varphi}
    =  e^{i(X_i Y_{i+1} - X_{i+1} Y_{i})\varphi/2}\\ 
    = \begin{pmatrix} 1&0&0&0\\ 0&\cos(\varphi)& -\sin(\varphi)&0\\
    0&\sin(\varphi)& \cos(\varphi)&0\\
    0&0&0&1 \end{pmatrix}
\end{multline}

\begin{figure}
    \centering
 \begin{quantikz}[align equals at=3.5,column sep=0.2cm, row sep={0.8cm,between origins}]
& \gate[wires=6]{\mathcal{U}_h^1(\theta)} & \qw \\
& & \qw \\
& & \qw \\
& & \qw \\
& & \qw \\
& & \qw 
\end{quantikz}
  =   \begin{quantikz}[align equals at= 3.5,column sep=0.2cm, row sep={0.8cm,between origins}]
& \gate[wires=2]{e^{-i(XX+YY)\theta/2}} & \qw \\
& & \qw \\
& \gate[wires=2]{e^{-i(XX+YY)\theta/2}}& \qw \\
& & \qw \\
& \gate[wires=2]{e^{-i(XX+YY)\theta/2}}& \qw \\
& & \qw \\
\end{quantikz}  
    \caption{Schematic of the quantum circuit $\mathcal{U}_h^1(\theta)$ for $N=6$ site chain for spin $\alpha$ (the same circuit will be applied to spin $\beta$ qubits indexed from $N$ to $2N-1$). A gate $e^{-i(XX+YY)\theta/2}$ is applied between each pair of qubits $2i$ and $2i+1$ for $i \in [0,N/2-1]$.}
    \label{fig:Uh1_scheme}
\end{figure}
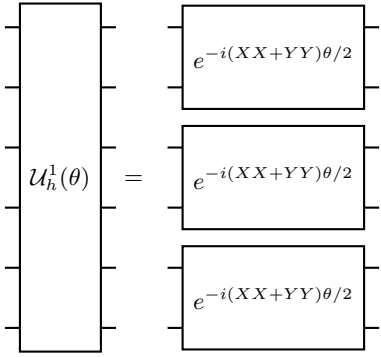

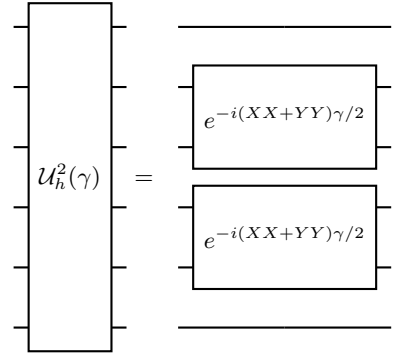
\begin{figure}
    \centering
  \begin{quantikz}[align equals at=3.5,column sep=0.2cm, row sep={0.8cm,between origins}]
& \gate[wires=6]{\mathcal{U}_h^2(\gamma)} & \qw \\
& & \qw \\
& & \qw \\
& & \qw \\
& & \qw \\
& & \qw 
\end{quantikz}
  =   \begin{quantikz}[align equals at= 3.5,column sep=0.2cm, row sep={0.8cm,between origins}]
&\qw&\qw\\
& \gate[wires=2]{e^{-i(XX+YY)\gamma/2}} & \qw \\
& & \qw \\
& \gate[wires=2]{e^{-i(XX+YY)\gamma/2}}& \qw \\
& & \qw \\
&\qw&\qw
\end{quantikz}  
   \caption{Schematic of the quantum circuit $\mathcal{U}_h^2(\gamma)$ for $N=6$ site chain for spin $\alpha$ (the same circuit will be applied to spin $\beta$ qubits indexed from $N$ to $2N-1$). A gate $e^{-i(XX+YY)\gamma/2}$ is applied between each pair of qubits $2i+1$ and $2i+2$ for $i \in [0,N/2-2]$.}
    \label{fig:Uh2_scheme}
\end{figure}

\begin{figure}
    \centering
    \begin{quantikz}[align equals at=1.5,column sep=0.3cm]
& \gate[wires=2]{G(\varphi)} & \qw \\
& & \qw 
    \end{quantikz}
   =  \begin{quantikz}[align equals at=1.5,column sep=0.3cm]
&\gate{H} &\ctrl{1}&\gate{R_y(-\varphi/2)}&\ctrl{1}& \gate{H} \\
&\qw &\targ{} & \gate{R_y(-\varphi/2)}&\targ{}& \qw
    \end{quantikz}
    \caption{Gate decomposition for Givens rotation.}
    \label{fig:Givens_decomp}
\end{figure}
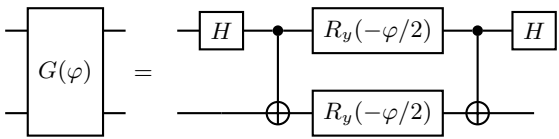

The last step is to find a gate decomposition that suits one's hardware specification. Aiming for universality, we choose in our simulations to decompose those unitary operations in terms of single-qubit gates and CNOTs. Figs. \ref{fig:Uh_decomp}, \ref{fig:Uu_decomp} show the decomposition for the unitary transformation associated to the hopping terms defined in Eq. \ref{eq:hopping} (proposed in Ref. \cite{gate_decomp}) and the interaction terms defined in Eq. \ref{eq:onsite}. Figs. \ref{fig:Uu_scheme}, \ref{fig:Uh1_scheme}, and \ref{fig:Uh2_scheme} show circuits associated to the different parts of the Hamiltonian. Fig. \ref{fig:Givens_decomp} displays the gate decomposition for Givens rotation (Eq. \ref{eq:givens}) taken from \cite{local_exp}. In this configuration, each element of the ansatz is decomposed with 2 CNOTs. For an N-site chain ($N > 2$) and a ladder qubit topology, this leads to 6N - 4 CNOTs per layer of the ansatz.

\begin{figure}
\centering
  \begin{quantikz}[align equals at=3.5,column sep=0.2cm, row sep={0.6cm,between origins}]
& \gate[wires=6]{\mathcal{U}_u(\phi)} & \qw \\
& & \qw \\
& & \qw \\
& & \qw \\
& & \qw \\
& & \qw 
\end{quantikz}
  =   \begin{quantikz}[align equals at= 3.5,column sep=0.2cm, row sep={0.6cm,between origins}]
&\ctrl{3}&\qw&\qw&\qw\rstick[wires=3]{$\uparrow$}\\
&\qw&\ctrl{3}&\qw&\qw\\
&\qw&\qw&\ctrl{3}&\qw\\
&\gate{e^{-iZZ\phi/2}}&\qw&\qw&\qw\rstick[wires=3]{$\downarrow$}\\
&\qw&\gate{e^{-iZZ\phi/2}}&\qw&\qw\\
&\qw&\qw&\gate{e^{-iZZ\phi/2}}&\qw\\
\end{quantikz}  
\caption{Schematic of the quantum circuit for $\mathcal{U}_u(\phi)$ for $N=3$ site Hubbard chain. A gate $e^{-iZZ\phi/2}$ acts between the qubit $i$ and $i+N$ for $i\in [0,N-1]$. }
\label{fig:Uu_scheme}
\end{figure}
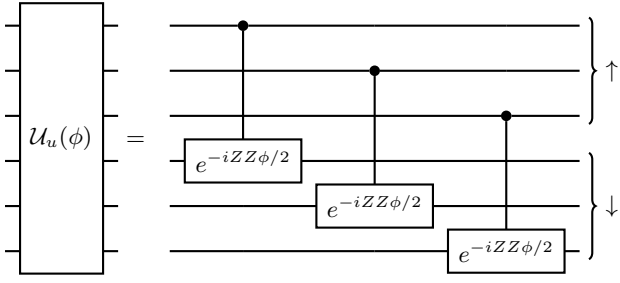

\begin{figure}
    \centering
    \begin{quantikz}[align equals at=1.5,column sep=0.3cm]
& \gate[wires=2]{e^{-i(X_i X_{i+1} + Y_i Y_{i+1})\theta/2}} & \qw \\
& & \qw 
    \end{quantikz}
   =  \begin{quantikz}[align equals at=1.5,column sep=0.3cm]
 &\gate{X}&\gate{R_x(\frac{\pi}{2})}&\ctrl{1}&
 \gate{R_x(\theta)}&\ctrl{1}&\gate{R_x(-\frac{\pi}{2})}& \gate{X}&\qw \\
 &\qw& \gate{R_x(-\frac{\pi}{2})}&\targ{}& \gate{R_z(\theta)} & \targ{}&\gate{R_x(\frac{\pi}{2})}&\qw  
    \end{quantikz}
    \caption{Gate decomposition for $e^{-i(X_i X_{i+1} + Y_i Y_{i+1})\theta/2}$.}
    \label{fig:Uh_decomp}
\end{figure}
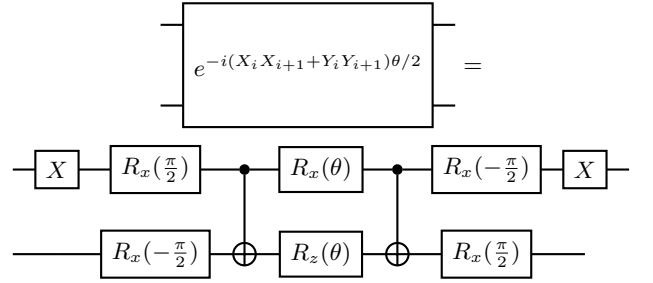
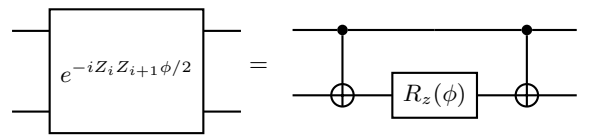
\begin{figure}
    \centering
    \begin{quantikz}
& \gate[wires=2]{e^{-iZ_i Z_{i+1}\phi/2}} & \qw \\
& & \qw 
    \end{quantikz}
   =  \begin{quantikz}
 &\ctrl{1}&\qw&\ctrl{1}& \qw \\
 &\targ{}& \gate{R_z(\phi)}&\targ{}& \qw
    \end{quantikz}
    \caption{Gate decomposition for $e^{-iZ_i Z_{i+1}\phi/2}$.}
    \label{fig:Uu_decomp}
\end{figure}

\subsection{Optimization details}\label{subsec:opt_details}

\begin{figure}[htb]
    \centering
    \includegraphics[width=0.75\linewidth]{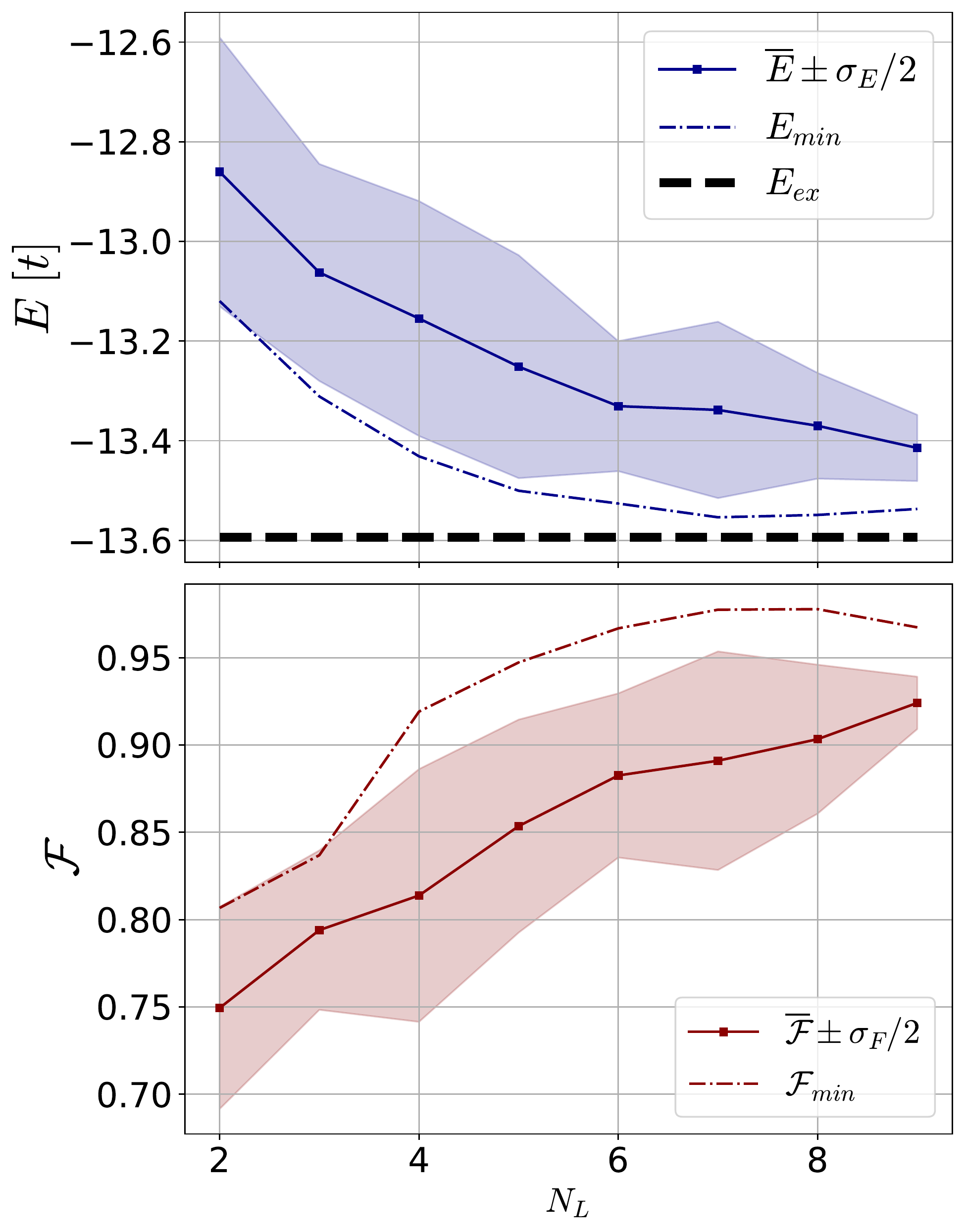}
    \caption{Averged optimized energy $E$ in units of $t$ over a set of 100 simulations with random initial parameters. The shaded areas correspond to the averaged energies (fidelities $\pm \sigma_E/2$ ($\pm \sigma_F/2$), $\sigma_i$ being the standard deviation. The colored dashed line corresponds to the minimal optimized energy and its associated fidelity.}\label{fig:multistart}
\end{figure}

We use a multi-start procedure for initialization simulation for $U_0 = 5$, which consists in trying different random initial parameters and keeping the lowest energy solution after minimization. To show how sensitive are the results with respect to the initial parameters, we try 100 random initial parameters contained in $[-\frac{\pi}{5}, \frac{\pi}{5}]$  and compare the obtained results. Fig. \ref{fig:multistart} shows the average optimized energies and their associated fidelities with standard deviation. 

\subsection{Noise model}\label{subsec:noise_model}

In order to model noise occurring in real devices, we use the Kraus operators representation to describe quantum channels associated with noise models (Eq. \ref{eq:kraus_channel}). We combine noise acting on idle qubits as well as gate errors. During their inactive periods, the qubits undergo simultaneously phase damping described by (Eq. \ref{eq:phase_damp}) which corresponds to the loss of quantum information without loss of energy and amplitude damping described by (Eq. \ref{eq:amp_damp}) which corresponds to spontaneous relaxation of qubits to $|0\rangle$.  
\begin{equation}\label{eq:kraus_channel}
    \rho' = \sum_k K_k \rho K_k^\dagger
\end{equation}

\begin{equation}\label{eq:phase_damp}
 K_0^{PD} = \sqrt{p^{PD}}\begin{pmatrix}
    1&0\\ 0&1
    \end{pmatrix}, 
     K_1^{PD} = \sqrt{1-p^{PD}}\begin{pmatrix}
    1&0\\ 0&-1
    \end{pmatrix}
\end{equation}
\begin{equation}\label{eq:amp_damp}
 K_0^{AD} = \begin{pmatrix}
    1&0\\ 0&\sqrt{1-p^{AD}}
    \end{pmatrix}, 
     K_1^{AD} = \begin{pmatrix}
    0&\sqrt{p^{AD}}\\ 0&0
    \end{pmatrix}
\end{equation}
The relaxation probability $p^{AD}$ is defined by typical time scale $T_1$ such that $p^{AD}(t) = 1 - e^{-t/T_1}$, $t$ being the idle time. Similarly, phase damping is associated to an error probability $p^{PD}$ with a typical time scale $T_\phi = (\frac{1}{T_2} - \frac{1}{2T_1})^{-1}$ such that $p^{PD}(t) = 1 - e^{-t/T_\phi}$ . 
Moreover, the gate imperfections are modelled by a depolarizing channel acting after each gate \cite{depol_rand}. The one-qubit channel is described by Eq. (\ref{eq:depol}) (the two-qubit version being the tensor product of Eq. (\ref{eq:depol})). As most benchmark specifications of quantum devices provides only errors rate for 1-qubit gate $\epsilon_1$ and 2-qubit gate $\epsilon_2$, we approximate the depolarizing probabilities by $p_1 = \epsilon_1$ and  $p_2 = \epsilon_2$. In Eq. (\ref{eq:depol}), $\{ X,Y,Z \}$ are the three Pauli matrices:
\begin{equation}\label{eq:depol}
    K_0^D = \sqrt{1-p}I, K_1^D = \sqrt{\frac{p}{3}}X,  K_2^D = \sqrt{\frac{p}{3}}Y,  K_3^D = \sqrt{\frac{p}{3}}Z.
\end{equation}
\subsection{Post-selection}\label{subsec:postselec}
\begin{figure}
    \centering
\begin{quantikz}
\qw&\ctrl{1}&\gate{H}& \ctrl{1}& \qw\\
\qw&\targ{}& \ctrl{-1} &\targ{}&\qw
\end{quantikz}
    \caption{Circuit performing the change of basis to measure $\frac{1}{2}(XX + YY)$.}
    \label{fig:circ_measure}
\end{figure}
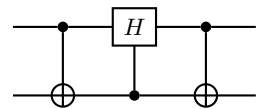

To mitigate errors coming from noise, we post-select the measurements that respect the number of particles and $S^z$ symmetries. To do this, we first need to choose a measurement scheme to evaluate the Hamiltonian. We use the following scheme proposed in \cite{strat_cade}:
\begin{itemize}
    \item The $ZZ$ terms are measured in the computational basis without need for change of basis circuit.
    \item To efficiently measure the hopping terms $\frac{1}{2}(XX + YY)$, we implement a change of basis $B$ that diagonalize the operator in the computational basis such that $\mathcal{B}^\dagger\frac{1}{2}(XX + YY)\mathcal{B} = D$, $D$ being diagonal. The circuit shown in Fig. \ref{fig:circ_measure} transforms a hopping term into $D = |01\rangle \langle01| -|10\rangle \langle10|$.
\end{itemize}
As $\mathcal{B}$ also conserves the number of particles for each spin, we can reject any shots that don't contain the right number of spin up and down electrons, as they are only produced by errors occurring in the device. For example, for a 2-site model, any measurement (when evaluating the hopping terms or the onsite interaction terms) that is not contained in $\{0101, 1010, 1001, 0110\}$ can be discarded. This method allows mitigating amplitude damping, read-out, and bit-flip errors.
\subsection{Richardson extrapolation}\label{subsec:richardson}
This error mitigation method was introduced first in Ref. \cite{error_mitig}. Let us suppose an open system composed of qubits initially prepared in $\rho_0$. We apply a quantum circuit described by a time-dependent driving Hamiltonian $K(t) = \sum_\alpha J_\alpha(t)P_\alpha$, where $P_\alpha$ are Pauli strings and $J_\alpha(t)$ real coefficients. The system will be described by Eq. (\ref{eq:rho_L}):
\begin{equation}\label{eq:rho_L}
    \frac{\partial}{\partial t}\rho(t) = -i[K(t), \rho(t)] + \lambda \mathcal{L}
(\rho(t)).
\end{equation}

$\mathcal{L}(\rho)$ is a noise generator that is invariant under time rescaling and independent from $K(t)$, which can be experimentally quite demanding. Then, an observable $\hat{O}$ can be measured on the state $\rho_\lambda(T)$ drived by $K$ for a duration $T$, such that $O_K(\lambda) = \text{tr}(\hat{O}\rho_\lambda(T))$, which can be expressed for $\lambda \ll 1$ ($\lambda$ being the experimental noise strength) by the following:
\begin{equation}
    O_K(\lambda) = O^* + \sum_{k=1}^n c_k \lambda^k + \mathcal{O}(\lambda^{n+1}),
\end{equation}
$O^*$ being the noise-free expectation value: $O^* = \text{tr}(\hat{O}\rho_0(T))$. Let assume one can scale up the noise level $\lambda$ by a factor $c$. By running the quantum circuit for different noise levels $\lambda_i = c_i\lambda$ for $i = 0,...,n$ ($c_0 =1, c_i > 1$ for $i>0$) one can obtain an improved estimate of $O^*$ up to precision $\mathcal{O}(\lambda^{n+1})$ expressed as:
\begin{equation}
    O^n_K = \sum_{i=0}^n\gamma_i O_K(\lambda_i),
\end{equation}
with the coefficients $\{\gamma_i\}$ being solutions of $\sum_i \gamma_i = 1$ and $\sum_i \gamma_i c_i^k = 0$ for $k = 1,...,n$. In the case $n = 1$, this is equivalent to a simple linear extrapolation:
\begin{equation}
    O^1_K = \frac{c_1-c_0}{c_1}(O_K(c_0\lambda) - O_K(c_1\lambda)) + O_K(c_1\lambda) 
\end{equation}

\nocite{*}

\bibliography{references}

\end{document}